\documentclass[twocolumn,prl,showpacs,preprintnumbers,amsmath,amssymb]{revtex4}

\usepackage{graphicx}
\usepackage{dcolumn}
\usepackage{bm}

\begin{document}

\title{
Tomonaga-Luttinger Liquid in a Quasi-One-Dimensional $S$=1 Antiferromagnet Observed by the Specific Heat }

\author{M. Hagiwara$^1$, H. Tsujii$^2$, C. R. Rotundu$^3$, B. Andraka$^3$,\\ Y. Takano$^3$, N. Tateiwa$^4$, T. C. Kobayashi$^5$, T. Suzuki$^6$, and S. Suga$^6$}
\affiliation{
$^1$KYOKUGEN, Osaka University, 1-3 Machikaneyama, Toyonaka, Osaka 
560-8531, Japan
\\ $^2$RIKEN (The Institute of Physical and Chemical Research), 2-1 Hirosawa, Wako, Saitama 351-0198, Japan \\ $^3$Department of Physics, University of Florida, P. O. Box 118440, Gainesville, Florida 32611-8440, USA \\
$^4$JAERI, Advanced Science Research Center, Tokai, Ibaraki 319-1195, Japan \\$^5$Department of Physics, Okayama University, 3-1-1 Tsushimanaka, Okayama 700-8530, Japan \\$^6$Department of Applied Physics, Osaka University, Suita, Osaka 565-0871, Japan}

\date{\today}

\begin{abstract}
Specific heat experiments on single crystals of the $S$=1
quasi-one-dimensional bond-alternating antiferromagnet
Ni(C$_9$H$_{24}$N$_4$)(NO$_2$)ClO$_4$, alias NTENP, have been performed in
magnetic fields applied both parallel and perpendicular to the spin chains. We have found for the parallel field configuration that the magnetic specific heat ($C_{\rm mag}$) is proportional to temperature ($T$) above a critical field $H_{\rm c}$, at which the energy gap vanishes, in a temperature region above that of the long-range ordered state. The ratio $C_{\rm mag}/T$ increases as the magnetic field approaches $H_{\rm c}$ from above. The data are in good quantitative agreement with the prediction of the $c$=1 conformal field theory in conjunction with the velocity of the excitations calculated by a numerical diagonalization, providing a conclusive evidence for a Tomonaga-Luttinger liquid.
\end{abstract}


\maketitle
%
%
One-dimensional antiferromagnets and their laboratory counterparts,
quasi-one-dimensional (quasi-1D) antiferromagnets, with an energy gap
above a singlet ground state are of current interest in condensed
matter physics. When an external magnetic field is applied to any of
these systems, one of the excited triplet branches goes down in
energy due to the Zeeman effect, and the energy gap vanishes at a
critical magnetic field $H_\mathrm{c}$. This is a quantum critical
point at which a transition from a gapped spin liquid to a
Tomonaga-Luttinger liquid (TLL) is expected to occur~\cite{sakai,
Sachdev}, if the system possesses a Heisenberg or XY symmetry. The
TLL is characterized by a gapless $k$-linear energy dispersion with
an incommensurate $k_0$ and a spin correlation having an
algebraic decay.

To date, however, all experimental evidence in favor of this
scenario has been either controversial or circumstantial. Divergence
of the NMR spin-lattice relaxation rate \cite{Chaboussant} and an
anomalous hump in the magnetic specific
heat \cite{Hammar,Calemczuk,HagiwaraCuHpCl} of the spin-1/2 ladder compound Cu$_2$(C$_5$H$_{12}$N$_2$)$_2$Cl$_{4}$, alias CuHpCl, at low temperatures 
have been interpreted to arise from a TLL.  A similar interpretation
of the NMR relaxation rate \cite{Izumi} and the magnetic specific
heat \cite{Yoshida} has been put forward for the spin-1/2
alternating-bond chain compound pentafulorophenyl nitronyl nitroxide
(F$_5$PNN). However, Stone \textit{et al.}~\cite{Stone} have
proposed on the basis of neutron measurements and an examination of
the crystal structure that CuHpCl is three-dimensional rather than
quasi-1D. Furthermore, neither compound exhibits a $T$-linear
specific heat indicative of the linear, gapless dispersion of a TLL.

The organometallic Ni-chain compound Ni$_2$(C$_5$H$_{14}$N$_2$)$_2$N$_3$(PF$_6$), abbreviated as NDMAP, is one of the best-established quasi-1D antiferromagnets with an XY symmetry. For this gapped $S$=1 linear-chain compound, the $H$-$T$ phase diagram of the field-induced long range order (LRO) in the magnetic fields above $H_\mathrm{c}$ for $H$$\parallel$chain has been studied in detail and has been explained by assuming that the
critical exponent of the TLL governs the LRO \cite{honda1}. However, no stronger evidence for a TLL exists. In this compound, the principal axis of the NiN$_6$ octahedra which determines the crystal-field anisotropy is tilted from the chain axis by 15$^{\circ}$, with the tilt direction alternating from chain to chain. Therefore, neither $H$$\parallel$chain nor other magnetic field directions strictly satisfy an XY symmetry for all the chains simultaneously. This lack of axial symmetry probably explains why the evidence for a TLL in NDMAP is only circumstantial at best.

A recently synthesized Ni compound Ni(C$_4$H$_{24}$N$_4$)NO$_2$(ClO$_4$), or NTENP~\cite{escuer} for short, is an $S$=1 bond-alternating-chain antiferromagnet. NTENP has an energy gap between the singlet ground state and the first excited triplet~\cite{narumi}, and exhibits an LRO above $H_{\rm c}$~\cite{tateiwa,hagiwara2}.  Unlike in NDMAP, the principal axis of the Ni$^{2+}$ ions in NTENP is nearly parallel to the chain direction. In this Letter, we report a TLL behavior observed in the specific heat of NTENP above $H_{\rm c}$. 

%
%
First, we briefly summarize the crystal and magnetic properties of NTENP by referring the reader to Ref.~\onlinecite{narumi} for details. NTENP belongs to a triclinic system (space group $P$$\bar{1}$), and the Ni$^{2+}$ ions are bridged along the $a$ axis by nitrito groups having two alternating bond lengths. The Ni chains are well separated by ClO$_4^-$ anions, thus having a good 1D nature. The inversion centers are situated in the nitrito groups rather than the Ni$^{2+}$ ions, so that no staggered components of the magnetic moments are expected to be retained. The critical fields $H_{\rm c}$ parallel and perpendicular to the chain direction are 9.3 T and 12.4 T, respectively~\cite{narumi}. 

The model spin Hamiltionian for NTENP is written as
 \begin{eqnarray}
   {\cal H} & = & \sum_{i} [J{\bf S}_{2i-1} \cdot {\bf S}_{2i}+\delta J{\bf S}_{2i} \cdot {\bf S}_{2i+1}-\mu_{\rm B}{\bf S}_i\tilde{g}{\bf H} \nonumber \\ &  & \mbox{}+D(S_i^z)^2+E\{(S_i^x)^2-(S_i^y)^2\}],
 \end{eqnarray}
where $J$ is the larger of the two exchange constants, $\delta$
the bond-alternating ratio, $\tilde{g}$ the $g$ tensor of
Ni$^{2+}$, $\mu_{\rm B}$ the Bohr magneton, and $D$ and $E$ the axial and orthorhombic single-ion anisotropy constants, respectively. From the analysis of the magnetic susceptibility data using numerical results with $E$=0, the following parameter values are known: $J/k_{\rm B}=54.2$~K, 
$\delta$=0.45, $D/J=0.25$, and $g_\|$=2.14 for the component of $\tilde{g}$ in the chain direction~\cite{narumi}. The ground state of NTENP is expected to be a singlet-dimer phase rather than a Haldane phase on the basis of the bond-alternating ratio. This expectation has been confirmed by magnetic susceptibility, magnetization, and ESR experiments on Zn-doped NTENP~\cite{narumi}. The analysis of the excitations observed by inelastic neutron scattering in the magnetic fields indicates that the strength of $E$ is only 5\% of that of $D$~\cite{hagiwara2}. For this reason, NTENP is an excellent candidate for an observation of a TLL. 
%
%

We prepared single crystals of hydrogenous and deuterated NTENP according to the method described in Ref.~\onlinecite{escuer}.
  The specific heat measurements for deuterated and hydrogenous NTENP in the magnetic fields applied along the chain direction ($a$ axis) were performed at the National High Magnetic Field Laboratory (NHMFL) in Tallahassee, Florida.  The measurements for a hydrogenous sample in the magnetic fields normal to the chain direction were performed at KYOKUGEN, Osaka University. Some of the results for this field configuration have been published in a preliminary report~\cite{tateiwa}. The lattice contribution of the specific heat was determined to be 3.31$\times$$T^3$ mJ/K mol by fitting the data at temperatures between 0.2 K and 1.1 K in zero field, where the singlet ground state with a large energy gap of about 1 meV~\cite{hagiwara2} makes the magnetic component of the specific heat negligible.  This lattice contribution has been subtracted from the raw data. In addition, we have subtracted the nuclear contribution of the hydrogen atoms from the hydrogenous-sample data at all the fields.

%
Figures 1(a) and (b) display the temperature dependences of the magnetic specific heat for deuterated and hydrogenous NTENP samples in the magnetic field applied parallel and perpendicular to the chains, respectively.   The hydrogenous sample for $H$$\parallel$chain has given similar results to those for the deuterated one. These results are not shown here, since they are less extensive than the deuterated-sample data.  For both field configurations, a sharp peak signals the LRO above the critical field, whose value depends on the field direction. The peak position moves up in temperature with an increasing magnetic field in both cases, but the behavior of the magnitude of the peak depends strongly on the field direction. For $H$$\parallel$chain, the peak height decreases slightly with an increasing magnetic field higher than 12 T as shown in Fig. 1(a), whereas it increases steeply with increasing field for $H$$\perp$chain (Fig. 1(b)).  

The most striking difference between the results for the two field directions is the temperature dependences in the region above the ordering temperatures. The specific heat for $H$$\parallel$chain is linear in temperature in this region, which becomes broad with increasing field. The specific heat is only weakly dependent on the magnetic field in this region. In contrast, a $T$-linear specific heat, which is indicative of a TLL state, is completely absent for $H$$\perp$chain. The specific heat above the temperature of the peak for a given magnetic field decreases much further with increasing temperature than for $H$$\parallel$chain and then gradually increases after reaching a broad, rounded minimum. A nearly field-independent behavior is found at temperatures further away from the peak. It is significant that a $T$-linear specific heat is not observed for $H$$\perp$chain. In this case, the breaking of the axial symmetry produces a crossover to a 2D Ising transition~\cite{affleck2}, causing an absence of a TLL.
\begin{figure}
\includegraphics[width=3in]{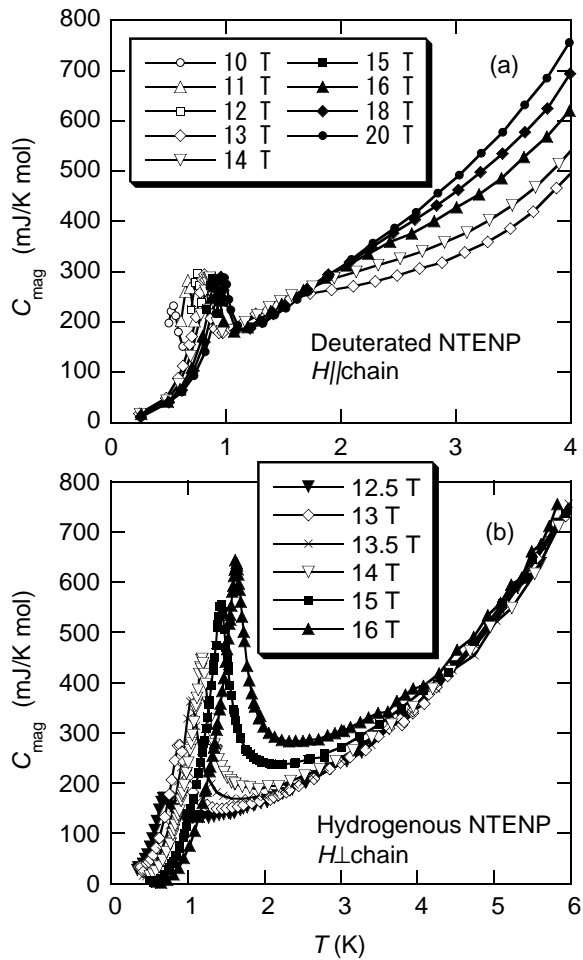}
\caption{(a) Magnetic specific heat of a deuterated sample for the magnetic fields applied parallel to the chains.
(b) Magnetic specific heat of a hydrogenous sample for the magnetic fields applied perpendicular to the chains.}
\label{fig:tdephc}
\end{figure}
\begin{figure}
\includegraphics[width=7.5cm,clip]{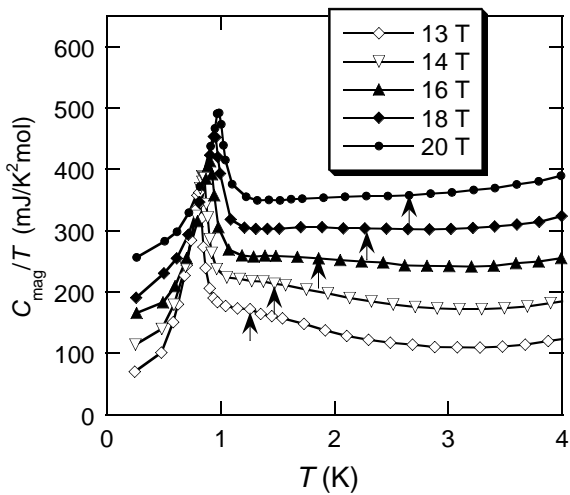}
\caption{Magnetic specific heat of the deuterated sample divided by temperature for the magnetic fields applied parallel to the chains.
For clarity, the data have been shifted up by 50 mJ/K$^2$mol at each field increment.} 
\label{fig:tdepcmag/t}
\end{figure}

To examine the linear temperature dependence in detail, we plot in Fig. 2 the specific heat for $H$$\parallel$chain divided by temperature for the five designated fields. We find that $C_{\rm mag}/T$ is nearly constant over a temperature range above the bottom of the high-temperature side of the peak. The upper end of the constant $C_{\rm mag}/T$ cannot be uniquely defined, since this is evidently a crossover point instead of a well-defined transition. We take an inflection point, which is marked by the arrow, to be the upper limit, although this choice is by no means the only possibility.  The upper boundaries of the TLL marked by the arrows in Fig. 2 are plotted in the $H$-$T$ phase diagram of Fig. 3. The diagram clearly indicates the stability of the TLL in high magnetic fields, in agreement with the theoretical expectation~\cite{chitra}.

\begin{figure}
\includegraphics[width=7.5cm,clip]{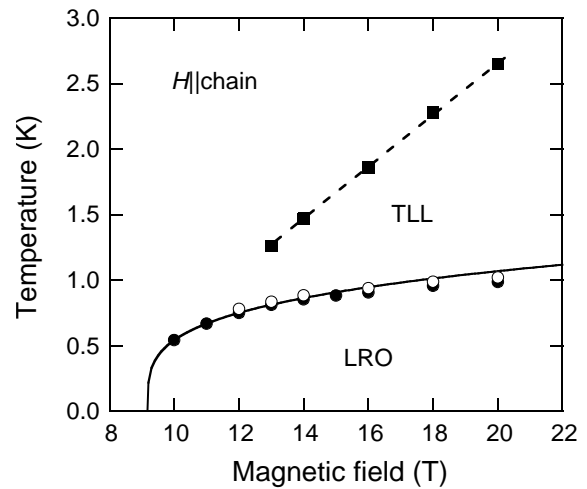}
\caption{Magnetic field versus temperature phase diagram of NTENP for the magnetic fields parallel to the chains. The regions marked LRO and TLL are the long-range ordered and Tomonaga-Luttinger liquid phases, respectively. The solid and open circles indicate the peak positions of the specific heat for the deuterated and hydrogenous samples, respectively, and the solid squares show the positions of the arrows in Fig. 2. The solid line is the best fit of the deuterated-sample data to the expression $T_{\rm c}=A|H-H_{\rm c}|^{\alpha}$ with $H_{\rm c}$=9.17 T and $\alpha$=0.264. The broken line is a guide to the eye.}
\label{fig:dispersion}
\end{figure}
%
%
The values of $\gamma$, defined as $C_{\rm mag}/T$ in the TLL region, are shown in Fig. 4 with solid circles. $\gamma$ is weakly field-dependent, rising slightly as the magnetic field decreases toward $H_{\rm c}\sim$9.3 T.  According to the conformal field theory for the central charge $c$=1~\cite{blote,affleck1}, $\gamma$ is given by $R\pi k_{\rm B}$/3$\hbar v$,  where $R$ and $v$ are the gas constant and the velocity of the low-lying excitations, respectively.  By using a numerical diagonalization method, we have calculated the dispersion curve of a chain of up to 20 spins and have extracted $v$ as a function of the ground-state magnetization. The results are sufficiently independent of the system size $N$ for $N$=16, 18, and 20. From these results and the experimental magnetization curve~\cite{narumi}, we obtain $\gamma$ as a function of the magnetic field. At fields up to about 16 T, the agreement with the experimental data is excellent as shown in Fig.~4. The calculated result at about 20 T is lower than the experimental value by about 20\%, but the agreement is still impressive. The increase of $\gamma$ with decreasing $H$ toward $H_{\rm c}$ is related to the divergence of the density of states at the band edge of the spinless fermions in a 1D system~\cite{usuki}.  The overall quantitative agreement between the experimental data and the calculation provides a conclusive evidence for a TLL.

\begin{figure}
\includegraphics[width=7.5cm,clip]{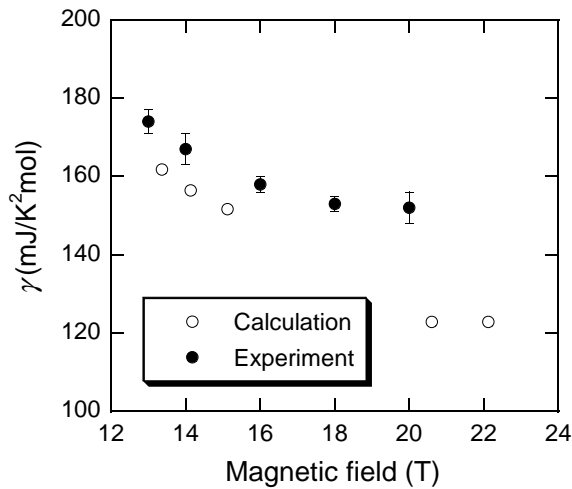}
\caption{Field dependence of $\gamma$=$C_{\rm mag}/T$. The solid circles are experimental values taken from Fig. 2, and the open circles are calculated ones, which are given by $R\pi k_{\rm B}/3\hbar v$.  Here, $R$ and $v$ are the gas constant and the velocity of the low-lying excitations, respectively.}
\end{figure}

The positions of the specific-heat peaks due to the LRO are shown in Fig. 3 for $H$$\parallel$chain.  The peak positions for the hydrogenous sample are also plotted for the same field direction. As observed in NDMAP~\cite{tsujii}, the ordering temperature is somewhat lower for the deuterated sample, indicating that deuteration slightly weakens the interchain exchange interaction. 

The LRO found above $H_{\rm c}$ is considered to be the Bose-Einstein condensation (BEC) of the triplets~\cite{nikuni}. According to the BEC theory~\cite{giamarchi}, the phase boundary in the $H$-$T$ phase diagram obeys the power law $T_{\rm c}\propto|H-H_{\rm c}|^{\alpha}$. The solid line is the best fit of the data for the deuterated sample up to 13 T to this expression with $H_{\rm c}$=9.17$\pm$0.01 and $\alpha$=0.264$\pm$0.002. The previously reported $\alpha$ of 0.334~\cite{tateiwa} was larger than this, because the transition occurred at higher temperature probably due to a tilting of the magnetic fields from the chain direction.  The present $T_{\rm c}$ values are consistent with those found by a recent NMR experiment~\cite{matsubara}. The $\alpha$ value of NTENP is smaller than 0.34 for NDMAP~\cite{tsujii}.   
  
 Finally, we comment that the field dependence of the peak height of the specific heat for $H$$\parallel$chain is unusual from a thermodynamic point of view.  Ordinarily, the peak at an LRO transition becomes larger as the critical region becomes wider with the increasing transition temperature due to an increasing magnetic field. However, the peak height observed for $H$$\parallel$chain decreases slightly with an increasing magnetic field.
The unusual field dependence probably arises from strong correlations in the TLL.  In the TLL, the correlation function decays only algebraically and the short range order is well developed. This leads to a substantial entropy drop above the LRO temperature. As the increasing magnetic field makes the TLL region wider, the entropy drop becomes larger, leaving less entropy to be expended at the LRO temperature. 
 
In conclusion, the specific heat of NTENP provides the first conclusive evidence for a TLL in a gapped quasi-1D antiferromagnet. We expect neutron scattering experiments of NTENP to observe a $k$-linear dispersion with an incommensurate $k_{\rm 0}$ in the TLL phase~\cite{hagiwara2}. But these experiments present technical challenges, because NTENP belongs to a triclinic system, for which the $a$ direction (the chain direction) and the direction of the reciprocal vector $a^{*}$ are different.


\emph{Acknowledgments---}
We thank A. Zheludev and D. L. Maslov for useful discussions, and G. E. Jones, T. P. Murphy, and E. C. Palm for assistance. This work was supported in part by the Grant-in-Aid for Scientific
Research on Priority Areas from the Japanese
Ministry of Education, Culture, Sports, Science and Technology, and by the NSF through DMR-0104240 and the NHMFL In-House Research Program. The NHMFL is supported by NSF Cooperative Agreement DMR-0084173 and by the State of Florida.


\end{document}